\documentclass[aps,prl,twocolumn,superscriptaddress]{revtex4}
\usepackage{amsmath,amssymb,graphicx,color,wasysym}
\usepackage{textcomp}
\usepackage{ulem}
\usepackage{color}

\definecolor{Phcolor}{rgb}{0.9,0,0.7}

\definecolor{GCcolor}{rgb}{0,0,0.9}
\definecolor{ADcolor}{rgb}{0,0.9,0}

\begin{document}

% Use the \preprint command to place your local institutional report
% number in the upper righthand corner of the title page in preprint mode.
% Multiple \preprint commands are allowed.
% Use the 'preprintnumbers' class option to override journal defaults
% to display numbers if necessary
%\preprint{}

%Title of paper
\title{Buckling instability causes  inertial thrust for spherical swimmers at all scales}

% repeat the \author .. \affiliation  etc. as needed
% \email, \thanks, \homepage, \altaffiliation all apply to the current
% author. Explanatory text should go in the []'s, actual e-mail
% address or url should go in the {}'s for \email and \homepage.
% Please use the appropriate macro foreach each type of information

% \affiliation command applies to all authors since the last
% \affiliation command. The \affiliation command should follow the
% other information
% \affiliation can be followed by \email, \homepage, \thanks as well.
\author{Adel Djellouli}
\affiliation{Univ. Grenoble Alpes, CNRS, LIPhy, 38000 Grenoble, France}
\author{Philippe Marmottant}
\affiliation{Univ. Grenoble Alpes, CNRS, LIPhy, 38000 Grenoble, France}
\author{Henda Djeridi}
\affiliation{Univ. Grenoble Alpes, Grenoble INP, CNRS, LEGI, 38000 Grenoble, France}
\author{Catherine Quilliet}
\affiliation{Univ. Grenoble Alpes, CNRS, LIPhy, 38000 Grenoble, France}
\author{Gwennou Coupier}
\affiliation{Univ. Grenoble Alpes, CNRS, LIPhy, 38000 Grenoble, France}
 \email{gwennou.coupier@univ-grenoble-alpes.fr}

\date{\today}

\begin{abstract} 

Microswimmers, and among them aspirant microrobots, generally have to cope with flows where viscous
forces are dominant, characterized by a low Reynolds number ($Re$). This implies constraints on the possible sequences of body motion, which have to be nonreciprocal. Furthermore, the presence of a strong drag limits the range of resulting velocities. Here, we propose a swimming mechanism, which uses the buckling instability triggered by pressure waves to propel a spherical, hollow shell. With a macroscopic experimental model,  we show that a net displacement is produced at all $Re$ regimes. An optimal displacement caused by non-trivial history effects is reached at intermediate $Re$. We show that, due to the fast activation induced by the instability, this regime is reachable by microscopic shells. The rapid dynamics would also allow high frequency excitation with standard traveling ultrasonic waves. Scale considerations predict a swimming velocity of order 1 cm/s for a remote-controlled microrobot, a suitable value for biological applications such as drug delivery.

\end{abstract}

% insert suggested PACS numbers in braces on next line
\pacs{}

\maketitle

Besides their playful aspect, artificial microswimmers present undeniable fundamental and practical interests, mostly driven by a constant race toward increasing miniaturization with potential applications such as targeted drug delivery. Comprehensive studies aim to identify the efficient strategies for small scale displacement in liquids \cite{stone96,alouges08,avron04,farutin13,ishimoto14,walker15,chisholm16}, which can possibly be exploited for the conception of synthetic microswimmers. Sticking to the strict definition of swimming as performing a displacement induced by body deformation, quite a few realizations of synthetic microswimmers can be found in literature \cite{najafi04,dreyfus05,peyer12,peyer13,ahmed16}. A growing attention toward the simplicity of their fabrication \cite{qiu15,bertin15,kaynak17} opens possibilities for transfer in the industrial arena. The two main external sources of power are magnetic \cite{peyer12,peyer13,qiu15} and acoustic  \cite{bertin15,ahmed16,kaynak17} fields, which are probably more suitable for medical applications and less expensive. The major conceptual difficulty lies in the low Reynolds flows usually associated with microscopic scales; the scallop theorem \cite{purcell77} then imposes that a non zero displacement may only occur via  a nonreciprocal succession of shapes. Except in chiral systems \cite{peyer12,peyer13}, this necessary condition requires at least two degrees of freedom, which commonly implies two control parameters. Such heavy double steering could indeed be bypassed if flow rates can be rendered high enough so that inertia cannot be neglected anymore, or if any hysteresis in the deformation "naturally" prevents reciprocity. 

\begin{figure}
\includegraphics[width=\columnwidth]{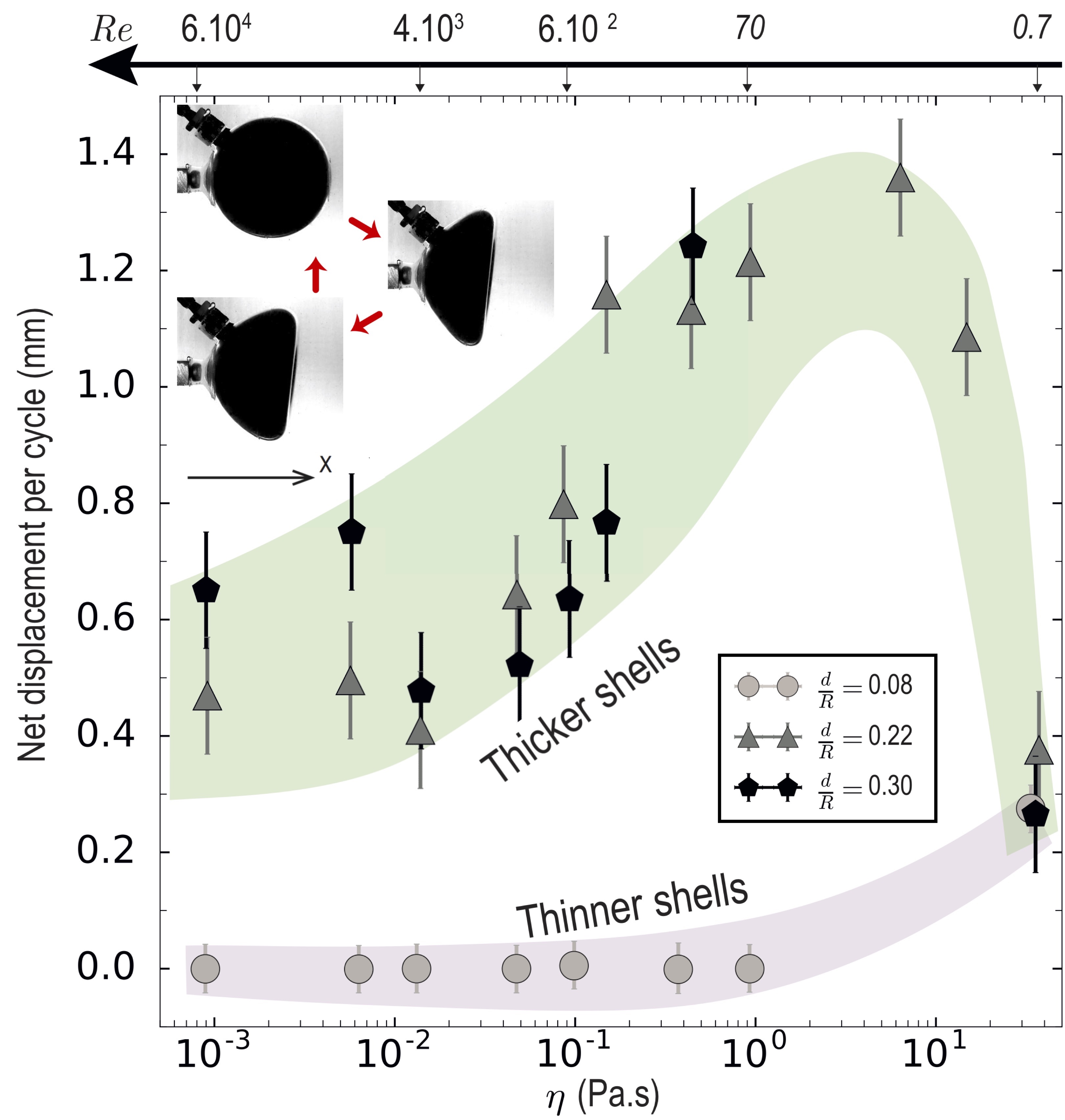}
\caption{\label{fig:displacement-tot} Inset: illustration of the deformation cycle. Main figure: displacement in the $x$ direction after one deformation cycle for three different $d/R$ ratios, as a function of fluid viscosity. The indicated Reynolds numbers correspond to $d/R=0.22$. Green (top) envelop highlights the high elastic energy regime while the low elastic energy regime is indicated by the pink (bottom) envelop.}
\end{figure}

 \begin{figure*}
\includegraphics[width=1.9\columnwidth]{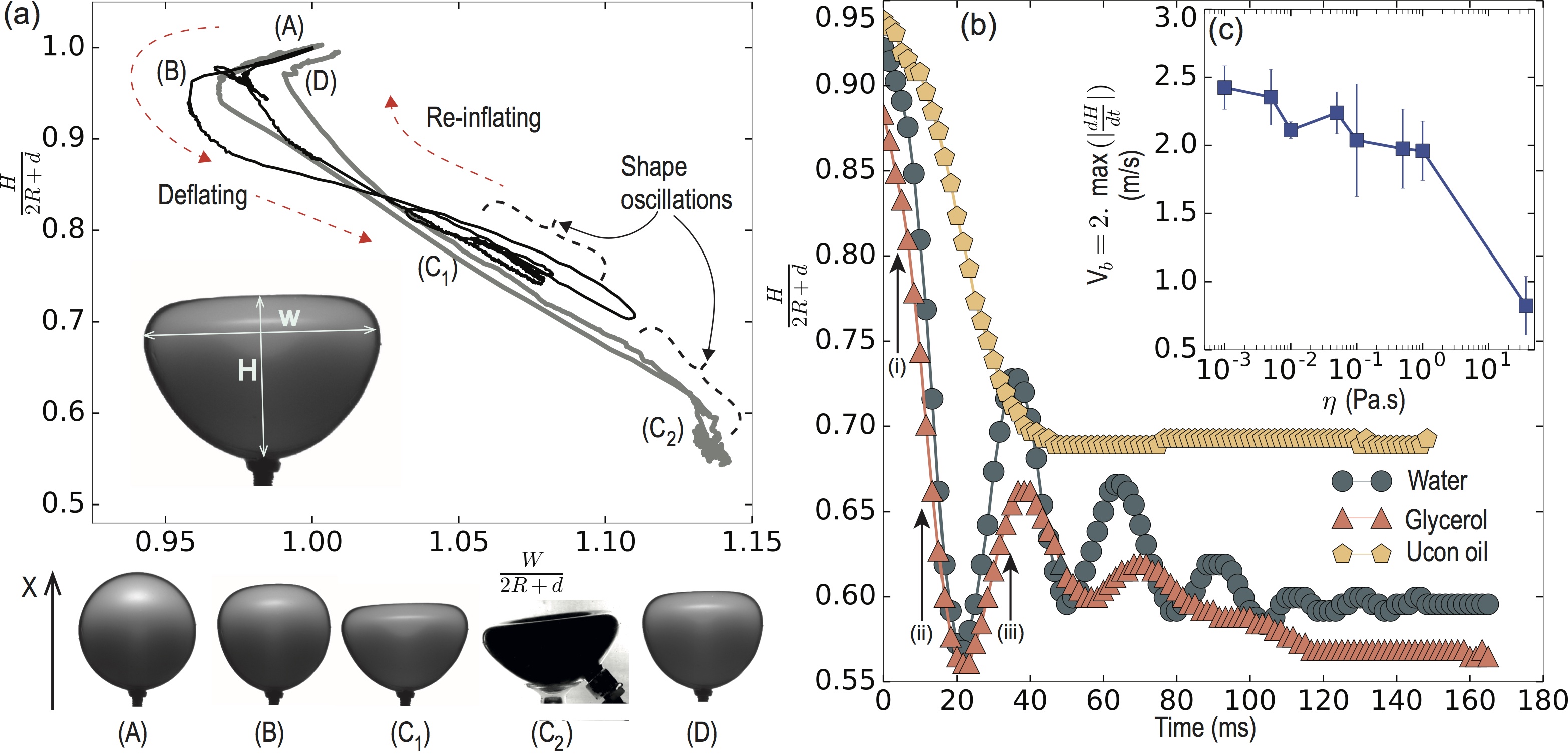}
\caption{\label{fig:deformation} Buckling and deformation of a shell with $d/R=0.22$. (a)  Path in a  height $H$ - width $W$  diagram along one pressure cycle in glycerol. Pressure difference $\Delta P$ is tuned by acting either on the internal pressure (thick gray curve), or on the external pressure (thin black curve). Pressure cycles of amplitude 0 to 600 mbar and period 15 s were applied. Pictures show the different shapes met by the shell along the pressure cycle: (A): initial spherical shape, (B): deformed shape right before buckling, (C\textsubscript{1}): buckled shape in the case of external pressure control - encapsulated air prevents full collapse, (C\textsubscript{2}): buckled shape in the case of internal pressure control, (D): shape after wrinkle unfolding right before unbuckling ; (b) Time evolution of the height $H/(2R+d)$ right after buckling (around  configuration  (C\textsubscript{2}) in Fig. a) in three different liquids: water ($\eta=10^{-3}$ Pa.s), glycerol ($\eta=0.9$ Pa.s), Ucon$^{\copyright}$ oil ($\eta=37$ Pa.s) ; (c) Buckling velocity $V_b$ as a function of fluid viscosity $\eta$.}
\end{figure*}

 We suggest fulfilling these two conditions together with simple spherical colloidal shells full of air that are microscopic objects quite easy to manufacture \cite{zoldesi05,pisani09}. Deflation from a spherical geometry occurs via buckling, which is a subcritical  instability, likely to provide both swiftness and hysteresis during a deflation-re-inflation cycle driven by a \textit{
single scalar control parameter}: pressure. We  investigate the swimming that results from these deformations thanks to macroscopic shells placed in a set of fluids with varying viscosity so that relevant dimensionless numbers could be kept unchanged from the microscopic scale. 
 
 \paragraph{Design and actuation of the swimmer} --- The swimmer was a hollow sphere of thickness $d$ and external radius $R+d/2=25$ mm, made in an elastomer  of Young modulus $E=0.5$ MPa. The pressure inside the shell was controlled by a pressure controller while the shell was attached to a frictionless rail. A weak spot for buckling was oriented in the rail direction ($x$ axis, inset of Fig. \ref{fig:displacement-tot}). Deformation without displacement was also studied  by controlling the external pressure so as to discuss the anticipated microscopic situation (activation by pressure waves). In that case, the shell was immersed in a pressurized tank (see \cite{supmat} for more details on the method).
 
 After a pressure cycle of sufficient amplitude so that buckling occurs, the shell and its support always move in the same direction whatever the shell thickness and the fluid viscosity (Fig. \ref{fig:displacement-tot}). Deeper insight into this swimming motion requires first to focus on the deformation dynamics.

\paragraph{Shell deformation cycle} --- A stress-free elastic spherical shell of radius $R$ and thickness $d$ submitted to an outside-inside pressure difference $\Delta P$ first shrinks while keeping its spherical symmetry, which corresponds to a quasi-linear relationship between $\Delta P$ and the volume variation (path A-B in Fig.~\ref{fig:deformation}-a)\cite{knoche11,marmottant11}. Then, over a threshold pressure difference $\Delta  P_{C}\simeq E(d/R)^2$ \cite{hutchinson67,landau86,knoche11,quilliet12}, an instability occurs toward a highly deflated conformation with a depression of extent $\sim R$ \cite{knoche11,quilliet12}.  In practice, the final state depends on the possibility to compress the inner medium (see e.g. C$_1$ and C$_2$ in Fig.~\ref{fig:deformation}-a). The depression often appears repeatedly on a weak spot, at a pressure difference possibly lower than $\Delta P_{C}$ \cite{lee16,zhang16}.  If $\Delta P$ is then decreased back to 0, another stable branch is followed, along which the radius of the depression decreases progressively much below $R$ (D in Fig.~\ref{fig:deformation}-a). Then, a small amplitude unbuckling instability brings the shell shape back to the isotropic branch.

\begin{figure}
\includegraphics[width=1.\columnwidth]{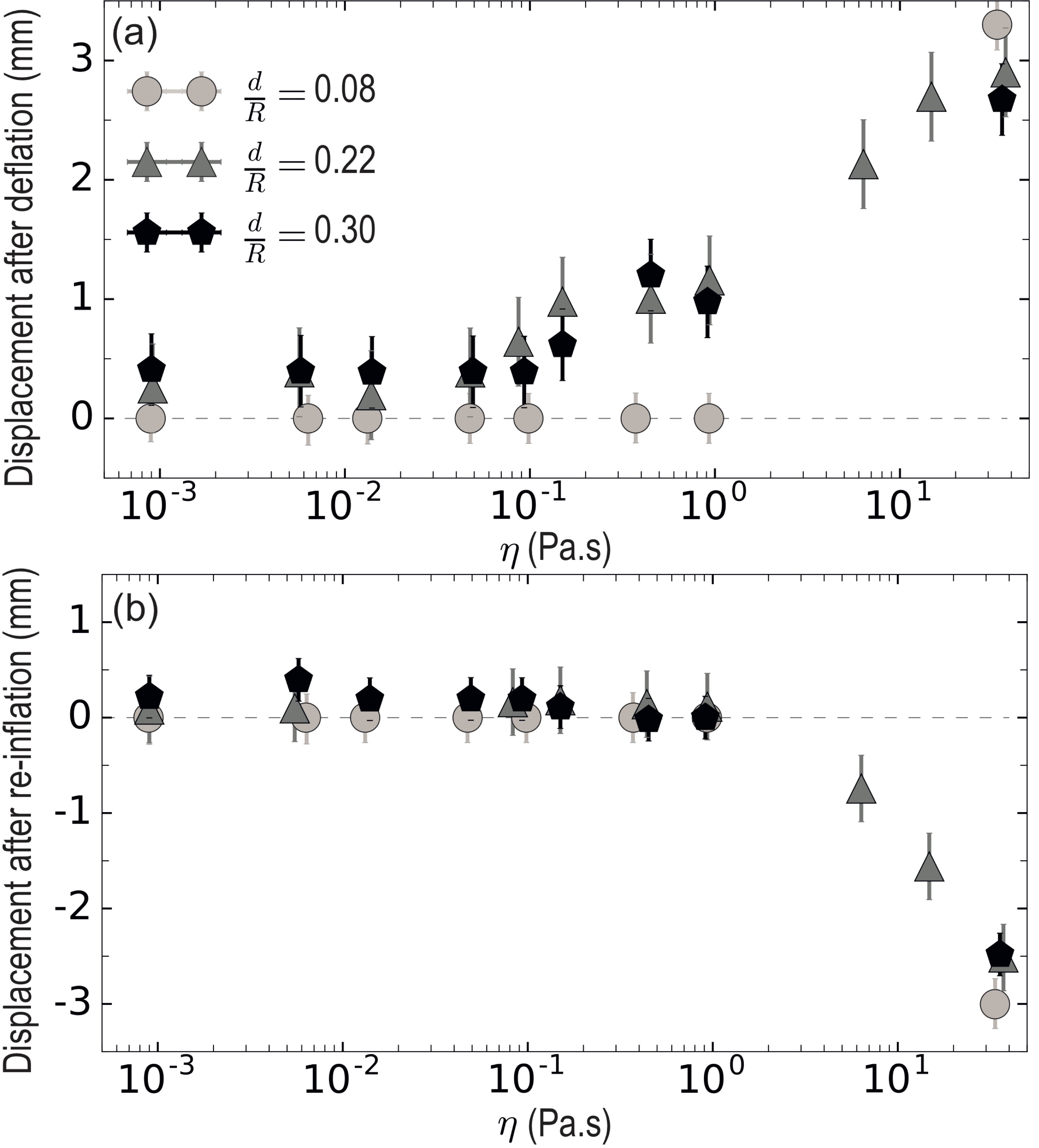}
\caption{\label{fig:displacement} Same as in Fig. \ref{fig:displacement-tot}, but decomposed into displacement during deflation and displacement during re-inflation.}
\end{figure}

\paragraph{Shape hysteresis and shape dynamics} --- The evolution in the  height $H$  and width $W$ of the shell during a pressure cycle exhibits the hysteresis that is a necessary condition for swimming at low Reynolds number (Fig.~\ref{fig:deformation}-a). 

After buckling, shape oscillations of frequency $\omega$ can be observed in most liquids  (Fig.~\ref{fig:deformation}-b). We define the buckling velocity $V_b$ as $2 \max(|dH/dt|)$, which should be close to the maximum velocity of the buckling spot. Fig.~\ref{fig:deformation}-c shows that this velocity is almost constant for fluid viscosities up to $\simeq$ 1 Pa.s after which the influence of fluid damping on the shell dynamics cannot be neglected. 

\paragraph{Displacements} --- Displacements as a function of fluid viscosity are shown in Fig.~\ref{fig:displacement}, for both phases of the cycle (increase or decrease of $\Delta P$), and three different relative thicknesses $d/R$.

 At low fluid Reynolds number (in Ucon oil of viscosity 37 Pa.s, $Re\equiv\rho_{f} V_b (R+d/2) /\eta =0.7$  for $d/R=0.22$), displacements are important in both phases (deflation and re-inflation), but they almost compensate within one cycle, with a final displacement of around 1\% of the radius due to shape hysteresis (Fig. \ref{fig:displacement-tot}).  In this Stokes regime, the displacement is quite similar for all shell thicknesses, as is the sequence of shapes.

On the opposite end of the viscosity range  ($Re=6.10^4$ in water), the inertial thrust should scale like $\rho_f \Delta \mathcal{V} \times  V_b \omega $, where $\rho \Delta \mathcal{V}$ is the mass of the accelerated fluid in the vicinity of the (un)buckling area and $V_b \omega$ its typical acceleration.

\begin{figure*}
\includegraphics[width=1.9\columnwidth]{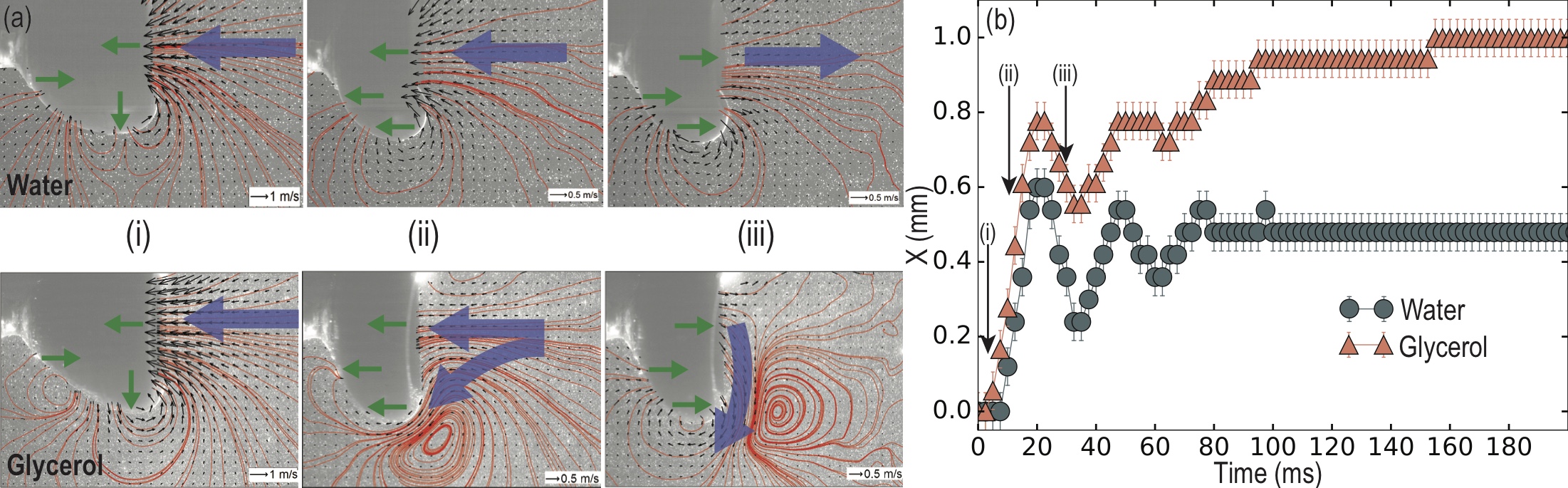}
\caption{\label{fig:flows} Post-buckling deformations and displacements for a shell with $d/R=0.22$. (a) Stream lines and velocity field during the shape oscillations, in water and in glycerol. Blue arrows indicate the main flow patterns. Arrows inside the ball indicate the main deformation direction ; steps (i) to (iii) are indicated on the deformation curve of Fig.~\ref{fig:deformation}-b. (i) and (ii) correspond to the initial inwards collapse and (iii) to the first outward oscillation ; (b) Displacement as a function of time.}
\end{figure*}

In our macroscopic model, this thrust serves first to accelerate the whole system (ball+gliding support) of mass $M\simeq 1.5$ kg. For the  $d/R=0.22$ shell, where  $\Delta \mathcal{V}$  is of order 20\% of the total volume $\mathcal{V}_0$ \cite{supmat}, this results in a typical swimming velocity $V_s=V_b\times 0.2 \rho_f \mathcal{V}_0/M\simeq 0.02 $ m/s, in very good agreement with the velocity measured during buckling (Fig. \ref{fig:flows}-b).

The accelerated volume $\Delta \mathcal{V}$ is an increasing function of $d/R$ \cite{supmat}, which makes the inertial regime eventually more efficient than the Stokes regimes  for thick enough shells (Fig. \ref{fig:displacement-tot}). Interestingly, this efficiency is also a consequence of the shape hysteresis: the pre-unbuckling shape (D in Fig.~\ref{fig:deformation}-a), which is obtained after a slow decrease of the depression, is much closer to the spherical shape than the post-buckling shape (C in Fig.~\ref{fig:deformation}-a). This leads to a much smaller amount of accelerated fluid at unbuckling, hence a negligible contribution of the re-inflation phase to the motion (Fig. \ref{fig:displacement}). For instance, for the $d/R=0.22$ shell,  the unbuckling volume change and velocity are smaller by a factor 8 and 2, respectively \cite{supmat}.

Last, the displacement after one full cycle appears to be even larger at an intermediate Reynolds number, for which the displacement due to buckling is enhanced. This points to the need for better knowledge of the surrounding fluid hydrodynamics. 

 \paragraph{ Flows} --- Flows around a $d/R=0.22$ shell attached to a fixed support were studied during buckling by time-resolved particle imaging velocimetry. They are qualitatively different according to the liquid viscosity (Fig.~\ref{fig:flows}-a), which echoes to distinct evolutions of the displacement during shape oscillations (Fig.~\ref{fig:flows}-b).

In water, the flow reverses during inward and outward shell oscillations, in phase with the boundary conditions. The buckling induces a displacement of the moving support that oscillates transiently in a synchronous way with the shape.

In glycerol, the not-fully-negligible viscous effect ($Re=70$) induces a qualitatively different scenario. The collapse of the shell during phase (ii) (backward motion of both back and front ends) creates a shear flow near the translating flank. This tangential flow is still present during the outward oscillation (iii). As a result, the outward motion of the expelled fluid is not directed towards the $x$ axis. This fluid does not contribute to the inertial thrust, which lowers the backward displacement (see phase (iii) in Fig.~\ref{fig:flows}-b). Thus, displacements due to oscillations do not counter each other as in water, but contribute, thanks to this delay effect, to a displacement lasting more than 100 ms (Fig.~\ref{fig:flows}-b).

The delay effect is characterized by the Womersley number $W\!o$,  defined as $W\!o^2=R^2 \frac{\rho_f \omega}{\eta}$, that compares the oscillation period with the viscous damping time. In the classical Stokes problem of a plate  oscillating with frequency $\omega$  in a viscous liquid, shear waves propagate in the normal $z$ direction with a wavenumber $k$ and a damping $e^{-k z}$, where $k=\sqrt{\rho_f \omega /( 2 \eta)}$  \cite{batchelor}. At a given distance $R$, the shear waves are damped at high $W\!o$ and they are in phase with the plate at low $W\!o$. For intermediate values of $W\!o\simeq 1$, the waves are nor damped nor synchronized. In our case, the phase shift gives rise to the complex pattern observed in glycerol where $W\!o\simeq 8$, while in water $W\!o\simeq 300$. Since $V_b \simeq R \omega$, $Re\simeq W\!o^2$ in our configuration ; thus, the interplay between inertial, non-stationary dynamics and viscous damping of shear waves  will always take place at an intermediate Reynolds number regime, and will imply an enhancement of the thrust during deflation.

\paragraph{Discussion and miniaturization} --- We discuss the implications of our results for the motion of a colloidal armored bubble in a water-like fluid, for which the control of the pressure difference would be attained by an external acoustic field.

\paragraph{Toy model} --- To that purpose, we developed a simplified model to describe the post-buckling dynamics of the shell \cite{supmatmodel}. If we consider a viscoelastic Voigt material of loss and storage modulus $E"(\omega)$ and $E'(\omega)$, with $E"(\omega)<E'(\omega)$ (which, in practice, is almost always the case \cite{lakes}), and if the surrounding fluid does not influence the shell dynamics, the post-buckling frequency obeys

\begin{equation}\label{eq:eqforWbis}  \omega \simeq \zeta   \times \frac{1}{R} \times\frac{d}{R}\times \sqrt{\frac{E'(\omega)}{\rho}},\end{equation}

with $0.3<\zeta<0.4$. Here, $\rho$ holds for the shell material volume mass. The buckling speed  $V_b$ obeys

\begin{equation}\label{eq:eqforV0}  V_b\simeq  \chi  \times\frac{d}{R}\times \sqrt{\frac{E'(\omega)}{\rho}},\end{equation}
with $0.4<\chi<0.9$. Note that by dimensionality arguments $V_b$ scales necessarily like $f(d/R) \times \sqrt{E'/\rho}$.

For our $d/R=0.22$ shell with $\rho=1060$kg/m$^3$, if we assume that $E'(\omega)=E'(0)\equiv E$, we find $\omega\simeq 64$ Hz, which, firstly is comparable the measured pulsation $\omega\simeq 150$ Hz  and secondly validates the above assumption since, for elastomeric materials, the stiffness $E'\left(\omega\right)$ is almost constant up to the kHz \cite{lakes}. We find that the buckling speed $V_b$ is comprised between 1.9 and 4.4 m/s, which perfectly surrounds the experimental value of $V_b\simeq 2.4$ m/s in water. 

This validates Eqs. \ref{eq:eqforWbis} and \ref{eq:eqforV0}, that allow to estimate the buckling velocities and post buckling frequency from the sole knowledge of $E'$ without that of $E"$.  This opens discussion for the possible scalings between microscopic systems and macroscopic ones.

\paragraph{ Flow regime} --- From the preceding calculation, we expect the Reynolds number to be  $Re\simeq 0.4 \frac{ R }{\eta}  (\rho\,\Delta P_C)^{1/2}$, with a prefactor $\simeq \sqrt{E'(\omega)/E}$. Displacement enhancement is controlled by the Womersley number and $W\!o\simeq Re^{1/2}$ in this problem.

%To identify the regime in which microswimmers would effectively lie, the accurate dimensionless numbers are the Reynolds number, the quality factor $Q_f=\omega \tau_f$, which has been shown to determine the shell dynamics in the fluid, and the Womersley number, which is crucial to understand swimming enhancement during shape oscillations. 

%Considering $\rho_f\simeq \rho$, the scalings are $Re\simeq 0.4 \frac{ R }{\eta}  (\rho\,\Delta P_C)^{1/2}$, $Q_f\simeq 0.1 \frac{ d }{\eta}  (\rho\,\Delta P_C)^{1/2}$,  with a prefactor $\simeq \sqrt{E'(\omega)/E}$. In addition, $W\!o^2\simeq Re$. 
%The scaling for $Re$ and $W\!o$ strictly holds when $Q_f>1$.

For a shell of radius $10$ \textmu m, and considering for $\Delta P_c$ the maximum value of 1 bar to avoid cavitation by the $\pm 1$ bar pressure wave, $Re \simeq 40 \sqrt{E'(\omega)/E}$. Eq. \ref{eq:eqforWbis} shows that, as $\omega$ scales as $R^{-1}$, miniaturization down to $10$ \textmu m propels $\omega$  to the MHz. Usual values of compliance for elastomers \cite{lakes} indicate that $Re$ may reach 400. Microscopic shells could then swim in the 1-600  intermediate $Re$ regime where inertial thrust is enhanced by the coupling between flow and shape oscillations. Besides, this happens at frequencies compatible with sonographic devices that are already known to induce repeated buckling on armored bubbles \cite{marmottant11}.

\paragraph{Expected displacements} --- A microscopic shell would be controlled by variations of the external pressure, while our macroscopic model was activated by varying the internal pressure.

At low Reynolds number, the sequence of shapes is quasi similar for both ways of controlling the pressure difference (Fig.~\ref{fig:deformation}-a) so we anticipate our result of a displacement per cycle of 1\% of $R$ (which is a slight underestimation due to friction on the arm holding the swimmer) to also hold for a microscopic shell.

At higher Reynolds number, the mass of accelerated fluid is given by the  loss of  shell volume  during buckling, which is limited by the resistance of the inner gas to compression. However, it can be shown that to the first two orders in $d/R$, the lost volumes are identical whatever the way the pressure is controlled \cite{supmat}.  The estimate done for the final displacement in the intermediate $Re$ (and $W\!o$) regime $1<Re<600$ is then valid, and is even a minor bound for a microswimmer that would not be attached to a heavy support.

%Second and third terms in Eq. \ref{eq:deltaV} denotes respectively  additional contribution to the thrust by inner gas compression and loss of thrust because of residual energy in the final buckled stage.

%From an energy perspective, we note that $E d^3$ is also  the stored elastic energy right before buckling \cite{supmat}.  At buckling, this elastic energy is then converted into work of $F_T$ on a typical distance $R$. 

%For slender bodies using undulatory gaits, equilibrating thrust and drag force allows to determine the swimming velocity in a permanent regime \cite{gazzola14}. Here the situation is more complex: the post-buckling oscillations and the buckling-unbuckling alternance lead to successions of positive and negative thrusts because of which body acceleration cannot be fully neglected. As indicated by our discussion on swimming at moderate Womersley numbers, the resulting actuation is a consequence of a subtle coupling of these oscillations with the fluid flow. Determining the full force-velocity relation requires further investigations. 

%Note that contrary to the Stokes regime this regime will be all the more efficient as the inertial thrust $F_T\propto R^{-1} \Delta \mathcal{V} \Delta P_C $ is high. This can be achieved, according to Eq.~\ref{eq:deltaV}, either by playing on $d/R$ or on $E$, within the limits $\Delta P_C<1 $bar and $d/R<1$. Furthermore, our results (Fig. \ref{fig:displacement}-c) show that a too large $d/R$ do not lead to higher efficiency, probably because of finite thickness effects that limit the collapse and the buckling velocity.
%

Finally, a microswimmer subjected to an ultrasonic wave of amplitude $\Delta P_C$ and driving frequency $\omega_d$ will swim under the condition that $\omega_d<\omega$, where $\omega$ is the post buckling spontaneous frequency of the shape oscillations, so as to allow time for the material to react to pressure variations. While the displacement per cycle is rather low, the high frequency that is allowed, thanks  to the fast activation due to the instability, may lead to high velocities.  We find that swimming velocity is at least equal to $U_s=0.01 R \times \omega/ (2 \pi)$ (the Stokes case). With $\omega_d\simeq \omega \simeq 1 $ MHz, this potentially leads to a net  velocity $\sim 1500 R$ per second, that is $15$ mm/s for a $R=10$ \textmu m shell, much faster than that of Janus particles (10 \textmu m/s \cite{paxton04}), helicoidal microrobots (10 \textmu m/s \cite{peyer12}),  microrobots with acoustically activated flagella (50 \textmu m/s \cite{ahmed16} to 1 mm/s but for much larger swimmers \cite{kaynak17}), microrobots  propelled by metachronal waves (3 \textmu m/s \cite{palagi13}).%, or living spherical microswimmers like Chlamydomonas Reinhardtii (50 \textmu m/s \cite{garcia11}). Recently, armored bubbles set into motion by acoustic streaming were found to swim at  1 mm/s \cite{bertin15}. 

A microscopic shell designed in such a way that the buckling pressure is of the order 1 bar would be in the intermediate Reynolds regime and would even swim faster. In addition, we anticipate that the swimming amplification observed in the moderate Womersley number regime opens a path for active amplification by a fine tuning of the pressure cycle period, so as to make it comparable to the viscous decay time.

Finally, we extrapolate that multidirectional remote control in a compound of several spheres of different characteristics, which may be built using smart self-assembly properties of colloidal particles \cite{yang08,cademartiri12,yi13},  can be reached by playing on the wave amplitude (with strong non-linear on/off response depending on whether the buckling pressure has been reached or not) and/or on the wave frequency. 

\paragraph{Conclusion} --- We have proposed and experimentally tested a swimming mechanism active at all $Re$ numbers, that relies on the intrinsic property of shape deformation hysteresis of a spherical shell upon a deflation/re-inflation cycle. Thanks to the fast deformations associated with shape instabilities, an inertial regime is reachable even at small scales. Hysteresis in the deformation velocity sequence implies fast propulsion in this regime, which can be amplified by the coupling between shape oscillations and flow patterns.

\begin{acknowledgments} This research has received funding from the European Research Council under the European Union's Seventh Framework Programme (FP7/2007-2013)/ERC Grant Agreement No.~614655 ÒBubbleboostÓ. We thank C.~Gr\'egoire for his contribution as an internship student, L.~Vignale and N.~Mordant for their help and advice for the PIV, T.~Combriat, P.~Peyla, S.~Rafa\"{\i} and T.~Podgorski for the numerous discussions, O.~Stephan and S.~Lecuyer for their help in the shell molding process, and all the workshop employees for their commitment to build up the different experimental setups. 
\end{acknowledgments}

%\subsection*{Author contributions}
%
%A.D., C.Q. and G.C. designed all the experiments but the PIV experiments.  A.D.  performed those experiments. A.D. and H.D. designed, performed and analyzed the PIV experiments. A.D., C.Q. and G.C.  analyzed data. G.C. developed the analytical model. P.M., C.Q. and G.C. conceived the project and supervised the experiments. A.D, C.Q. and G.C wrote the paper. All authors contributed to the paper proof readings and improvement discussions.

%\subsection*{Additional information}
%
%Supplementary Information accompanies this paper.
%
%Competing financial interests: The authors declare no competing financial interests.
%
%Correspondence and requests for materials should be addressed to gwennou.coupier@univ-grenoble-alpes.fr
\end{document}